\documentclass[letter]{aa}

\usepackage{graphicx}
\usepackage{pdflscape}
\usepackage{booktabs}
\usepackage{txfonts}

\begin{document} 

   \title{Rotational splitting as a function of mode frequency for six Sun-like stars \thanks{Table 2 is only available at the CDS via anonymous ftp to cdsarc.u-strasbg.fr (130.79.128.5) or via http://cdsarc.u-strasbg.fr/viz-bin/qcat?J/A+A/568/L12}}

   \author{M. B. Nielsen
          \inst{1,2}
          \and
	        L. Gizon
          \inst{2,1}
          \and
          H. Schunker
          \inst{2}
          \and
          J. Schou
          \inst{2}
          }

   \institute{Institut f{\"u}r Astrophysik, Georg-August-Universit{\"a}t G{\"o}ttingen, Friedrich-Hund-Platz 1, 37077 G{\"o}ttingen, Germany\\
             \email{nielsenm@mps.mpg.de}
             \and
	 			 Max-Planck-Institut f{\"u}r Sonnensystemforschung, Justus-von-Liebig-Weg 3, 37077 G{\"o}ttingen, Germany
             }

   \date{Received 3 July 2014; Accepted 31 July 2014}

  \abstract
	{Asteroseismology offers the prospect of constraining differential rotation in Sun-like stars. Here we have identified six high signal-to-noise main-sequence Sun-like stars in the \textit{Kepler} field, which all have visible signs of rotational splitting of their $p$-mode frequencies. For each star, we extract the rotational frequency splitting and inclination angle from separate mode sets (adjacent modes with $l=2$, $0$, and $1$) spanning the $p$-mode envelope. We use a Markov chain Monte Carlo method to obtain the best fit and errors associated with each parameter. We are able to make independent measurements of rotational splittings of $\sim8$ radial orders for each star. For all six stars, the measured splittings are consistent with uniform rotation, allowing us to exclude large radial differential rotation. This work opens the possibility of constraining internal rotation of Sun-like stars.}
	
   \titlerunning{Rotational frequency splitting in Sun-like stars}
   \authorrunning{Nielsen et al.}
   \maketitle

\section{Introduction}
Until recently, measuring the rotation of a star other than the Sun has been restricted to measuring the rotation rate at, or near, the photosphere. Techniques such as spectral line broadening obtain the projected rotational velocity $v \sin i$ \citep{Kraft1970,Gray2005}, where $i$ is the inclination of the stellar rotation axis with respect to the line of sight. However, this is difficult for slowly rotating stars \citep[e.g.,][]{Reiners2012} and a measurement is fundamentally ambiguous because of the often unknown inclination. An alternative to this approach is to analyze photometric light curves for signs of active regions crossing the stellar disk \citep{Nielsen2013,Reinhold2013,McQuillan2014}. If enough crossing events of sufficient contrast and coherence are seen, one can estimate the rotation period. However, for a star like the Sun this is not always possible because the short lifetime of active regions \citep{Solanki2003} compared to the mean solar rotation period leads to an incoherent signature in integrated light.

Asteroseismology is a tool that can be used to independently measure stellar rotation. In stars like the Sun the outer convective zone randomly excites acoustic oscillations (called $p$-modes) that propagate through the stellar interior. These oscillation modes can be described by a set of spherical harmonic functions of angular degree $l$ and azimuthal order $m$, as well as radial order $n$. Modes with $|m|>0$ travel around the rotation axis of the star in prograde and retrograde motion. For a non-rotating star the frequencies of these modes are degenerate with that of the $m=0$ modes, but become Doppler shifted if the star is rotating. This frequency shift, or splitting, is linearly related to the rotation rate of the star \citep[see, e.g.,][for details]{Aerts2010}. This effect has been exploited to image the internal rotation in the Sun \citep[see, e.g.,][]{Schou1998}. 

Using the high-quality observations of stellar light curves from space borne missions such as CoRoT \citep{Fridlund2006} and \textit{Kepler} \citep{Borucki2010}, it is possible to detect this frequency shift of the azimuthal modes in stars. This was done for a sample of subgiant stars by \citet{Deheuvels2012,Deheuvels2014}, for which it is possible to measure radial differential rotation because of the presence of mixed modes. These modes are sensitive to conditions in both the core and the outer envelope, thereby revealing the rotation rate at different depths in the star. This has been achieved for a wide variety of stars such as pulsating B-type stars \citep[e.g.,][]{Aerts2003,Pamyatnykh2004}, white dwarfs \citep{Charpinet2009}, and a main-sequence A-star \citep{Kurtz2014}. However, stars like the Sun only exhibit pure acoustic modes, which are primarily sensitive to conditions in the outer envelope, and so measurements are dominated by the rotation rates in this part of the star. The average rotation for a few Sun-like stars has been measured using data from CoRoT \citep{Gizon2013} and \textit{Kepler} \citep{VanEylen2014,Lund2014}. In this paper we perform an asteroseismic analysis of six Sun-like main-sequence stars observed by \textit{Kepler}, and measure the rotational splittings from their oscillation spectra.

\section{Analysis}
The splitting of oscillation modes by rotation in Sun-like stars is typically only seen as a broadening rather than a distinct separation of the modes, caused by the combined effect of the mode linewidths and the slow rotation rates. We handpicked Sun-like stars with the longest observed time series to get the highest possible frequency resolution, and a high signal-to-noise ratio near the $p$-mode envelope. The typical length of a time series used in this work spans $\sim3$ years. We defined these stars as Sun-like based on their reported temperatures $T_{\rm{eff}} \sim 6000$K and surface gravities $\log{g} \gtrsim 4$ from spectroscopic measurements by \citet{Bruntt2012} and \citet{Zakowicz2013}. We found six stars with these characteristics that also have visible rotational splitting of the $l=1$ and $l=2$ modes. 

We used the Lomb-Scargle method as applied by \citet{Frandsen1995} to compute the power spectrum of each light curve. We fit a model of the oscillation modes to small segments of the power spectrum spanning a set of $l=2,0,1$. These mode sets are all consecutive in frequency and together span the $p$-mode envelope of the star in question (see Table \ref{tab:results}). We used maximum likelihood estimation to find the best-fit solution and obtain a rotational splitting for each mode set. 

\subsection{Observations}
We used short-cadence ($\sim 58$ second) white light observations from the NASA \textit{Kepler} mission from March 2009 until the end of the mission in early 2013. The data were obtained from the Mikulski Archive for Space Telescopes\footnote{http://archive.stsci.edu/kepler/}. We used data that was pre-processed by the PDC\_MAP and msMAP pipelines \citep{Smith2012,Thompson2013} prior to release. In some cases we found narrow peaks caused by residual instrumental effects \citep{Christiansen2011}, although none of these overlapped with the $p$-mode frequencies. However, we note that the various background noise terms which we included in our model could potentially be influenced by the presence of instrumental peaks. 

The \textit{Kepler} Input Catalogue (KIC) numbers for the analyzed stars are shown in Table \ref{tab:results}, along with the spectroscopic effective surface temperature and surface gravity measurements \citep{Bruntt2012,Zakowicz2013}. 

\subsection{Power spectrum model}
We fit the power spectrum with a model consisting of a constant noise level, two frequency-dependent Harvey-like noise terms \citep[see Equation 1 in][]{Aigrain2004}, in addition to the individual oscillation modes. We model these as a sum of Lorentzian profiles as per Equation~10 in \citet{Handberg2011}, each consisting of mode power, frequency, and linewidth. 

We perform an initial fit of the background noise components to the entire spectrum of each star. These background terms are caused by various processes in the stellar photosphere such as granulation and magnetic activity, and span a wide range in frequency that often overlaps with the $p$-mode oscillations. We found that using two background terms was sufficient to account for the noise down to $\sim 10-100\mu$Hz, while the $p$-mode oscillations of the stars considered here have frequencies $>1000\mu$Hz. The fit parameters describing the background are subsequently kept fixed when fitting the $p$-mode oscillations.

\begin{figure}
	\centering
		\includegraphics[width = 0.90\columnwidth]{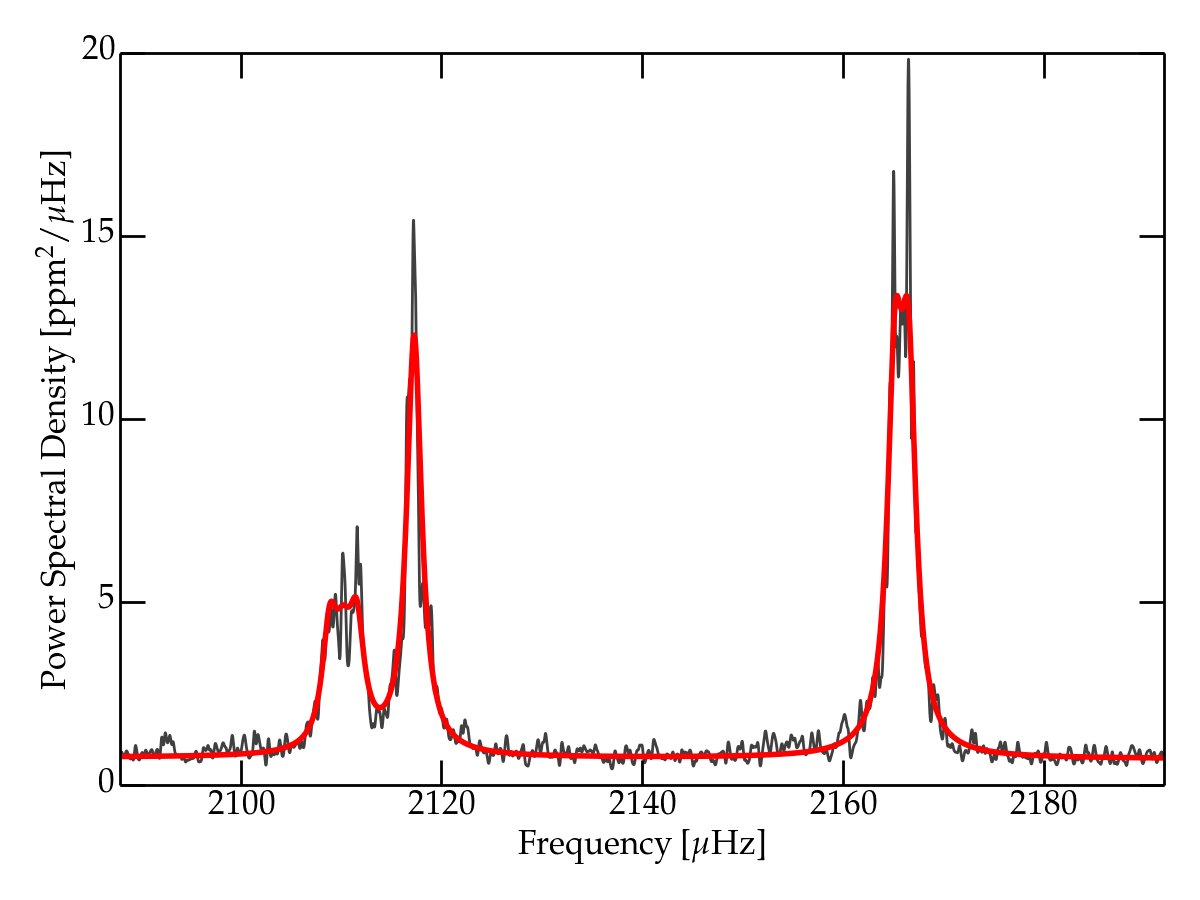}
		\caption{Example of a local fit performed to a segment ($6th$ modeset) of the power spectrum of \object{KIC006106415}. The power spectrum smoothed with $0.1\mu$Hz wide Gaussian kernel is shown in black. The red curve shows the best-fit model.}
	\label{fig:perky_modelfig}
\end{figure}

We divide the p-mode envelope into segments of length roughly equal to the separation between radial orders (called the large frequency separation), and centered approximately between the $l=0$ and $l=1$ modes. Thus a segment contains a set of modes of angular degree $l=2$, 0, and 1 (see Fig.~\ref{fig:perky_modelfig}), which we fit separately from the other sets in the spectrum. 

For each angular degree $l$ there is multiplet of $2l+1$ azimuthal modes, where, for a slowly rotating star, the components are mutually separated by an amount proportional to the stellar rotation rate $\Omega$. The frequencies of these modes can be expressed as
\begin{equation}
\nu _{nlm}  = \nu _{nl}  + m\frac{\Omega}{{2\pi }}\left( {1 - C_{nl} } \right) \approx \nu _{nl}  + m\delta \nu,  
\end{equation}
where $\nu _{nl}$ is the frequency of the central $m=0$ mode, with the $|m|>0$ modes displaced from this frequency by the effect of rotation. The value $C_{nl}$ is small for modes of $n  \gtrsim 20$ in Sun-like stars and is considered negligible. We can therefore approximate the frequency shift of the azimuthal orders by the amount $m \delta \nu$, where the rotational splitting $\delta \nu$ is equivalent to the rotation frequency of the star. We assume a common rotational splitting for the modes of a given mode set.

The amplitude of the Lorentzian profiles is a product of the mode height and the mode visibility. The mode heights are left as free parameters, and are assumed equal for all the components of a given $l$. The mode visibility is a function of the inclination $i$ of the star, where we fit a common inclination for the modes of each set. We use the form of the mode visibility as in \citet{Gizon2003}.

For stars that rotate pole-on relative to our point of view ($i=0$), the visibility of the $|m|>0$ modes approach zero, and so rotation cannot be measured. However, if a splitting of the $l = 1$ or $l=2$ modes can be measured the different visibilities of the modes allow us to infer the inclination of the stellar rotation axis, which is not easily done using other types of observations like spectroscopy, for example. 

Each mode has a finite width proportional to the lifetime of the oscillations, which is typically only on the order of a few days for Sun-like stars. For slowly rotating stars the rotational splitting may be small compared to the broadening caused by the lifetime of the mode. This makes it difficult to identify the individual azimuthal orders. However, the $l=0$ mode is unaffected by rotation so the linewidth of this mode can be taken as representative of the $l=1$ and $l=2$ modes and their associated azimuthal components \citep{Chaplin1998}. Thus, for each set of $l=2,0,1$ modes we assume a common mode linewidth.

\subsection{Fitting}
\begin{figure*}
	\centering
		\includegraphics[width = 1.93\columnwidth]{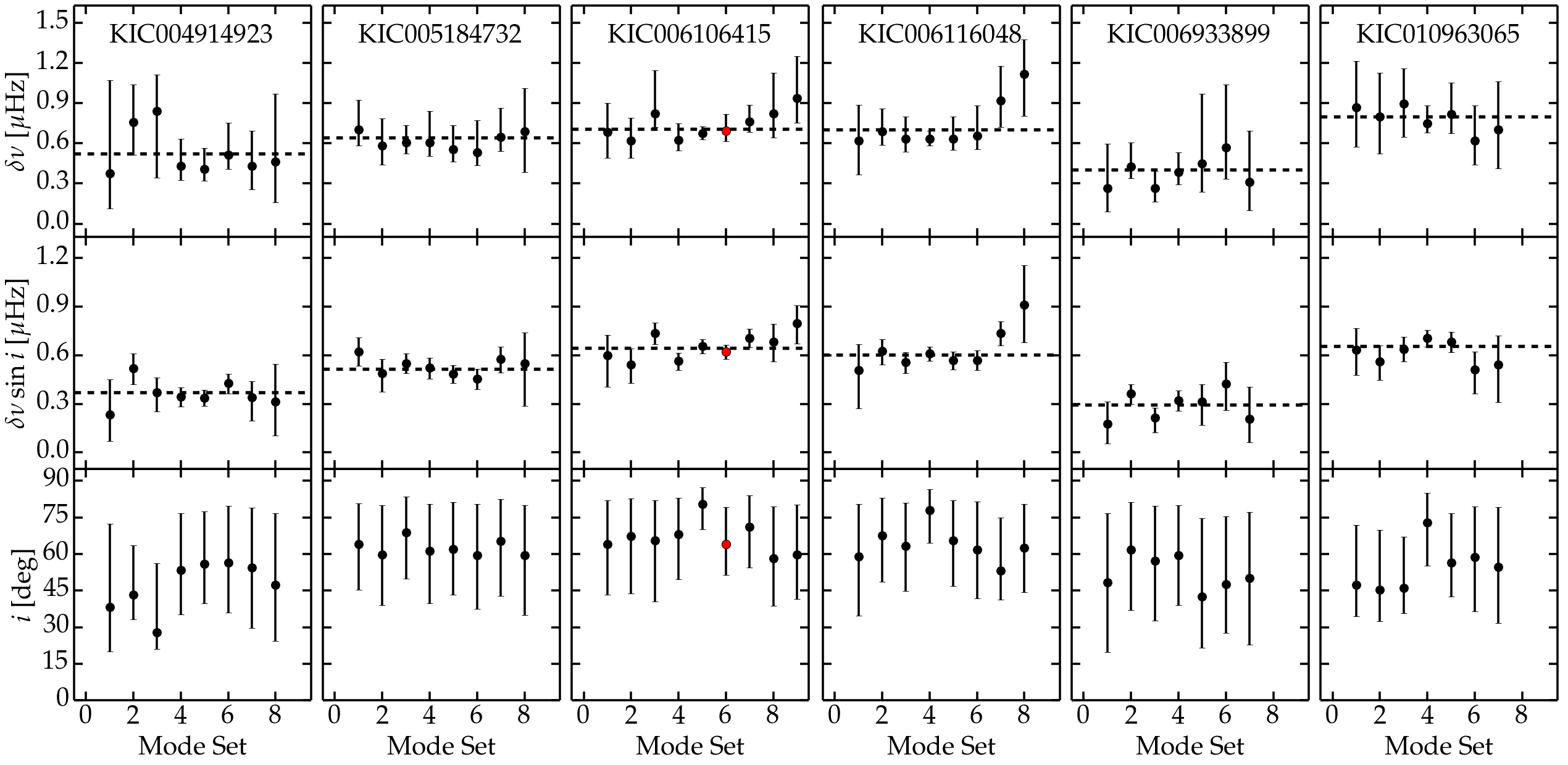}
		\caption{The measured splitting $\delta \nu$ and inclination $i$ of each mode set. The points show the results of the local fit as a function of the mode sets in each power spectrum. The error bars denote the $16th$ and $84th$ percentile values of the marginalized posterior distributions obtained from the MCMC samples. Dashed lines indicate the variance weighted mean of the values, using the variance of each posterior distribution. Red points show the mode set used in Fig. \ref{fig:perky_modelfig}. }
	\label{fig:split_and_incl}
\end{figure*}

We used a Markov chain Monte Carlo (MCMC) sampler\footnote{The affine invariant sampler in the EMCEE package for Python, http://dan.iel.fm/emcee/current/} \citep{EMCEE2013} to find the best-fit solution. The likelihood was computed using a $\chi^2$ probability density function as in \citet{Anderson1990}. We used the MCMC chains to compute the marginalized posterior distributions for each parameter, where we adopt the median of each distribution as a robust measure of the best-fit parameter value. We estimated the lower and upper errors for each parameter by the $16th$ and $84th$ percentile values of the posterior distributions. Figure \ref{fig:perky_modelfig} shows a model fit to a section of the power spectrum of \object{KIC006106415}. 

For the mode heights, central frequencies and width the initial positions of the walkers were randomly chosen from a normal distribution centered on a manually-determined initial guess. Each distribution had a standard deviation equal to $10\%$ of the initial guess value in order to provide the walkers with sufficient initial coverage of parameter space. For the inclination and splitting parameters we opted to use a uniform random distribution between $0-90^{\rm{o}}$ and $0-2\mu$Hz, respectively, since these two parameters are known to be non-linearly correlated. 

We used a probability distribution function of $\sin i$ on the inclination angle as a prior. We used uniform priors for all other parameters. These were only constrained for the $l=2$ and $l=0$ frequencies and the rotational splitting, such that the frequencies of each mode could not overlap. Initial testing showed that the walkers of the MCMC chain would sometimes switch the frequencies of these two modes because of their proximity. We found that this limitation on $\delta \nu$ did not bias the measurements or errors after inspection of the posterior distributions. 

We used 100 walkers to generate the MCMC chains which were allowed to run for $1200$ steps, giving us 120\,000 samples in the available parameter space. Although not strictly necessary owing to the rapid mixing of the walkers, we chose to disregard the first 600 steps as the burn-in phase of the MCMC chains. 

\section{Rotation and inclination as a function of frequency}
\begin{figure}
	\centering
		\includegraphics[width = 0.99\columnwidth]{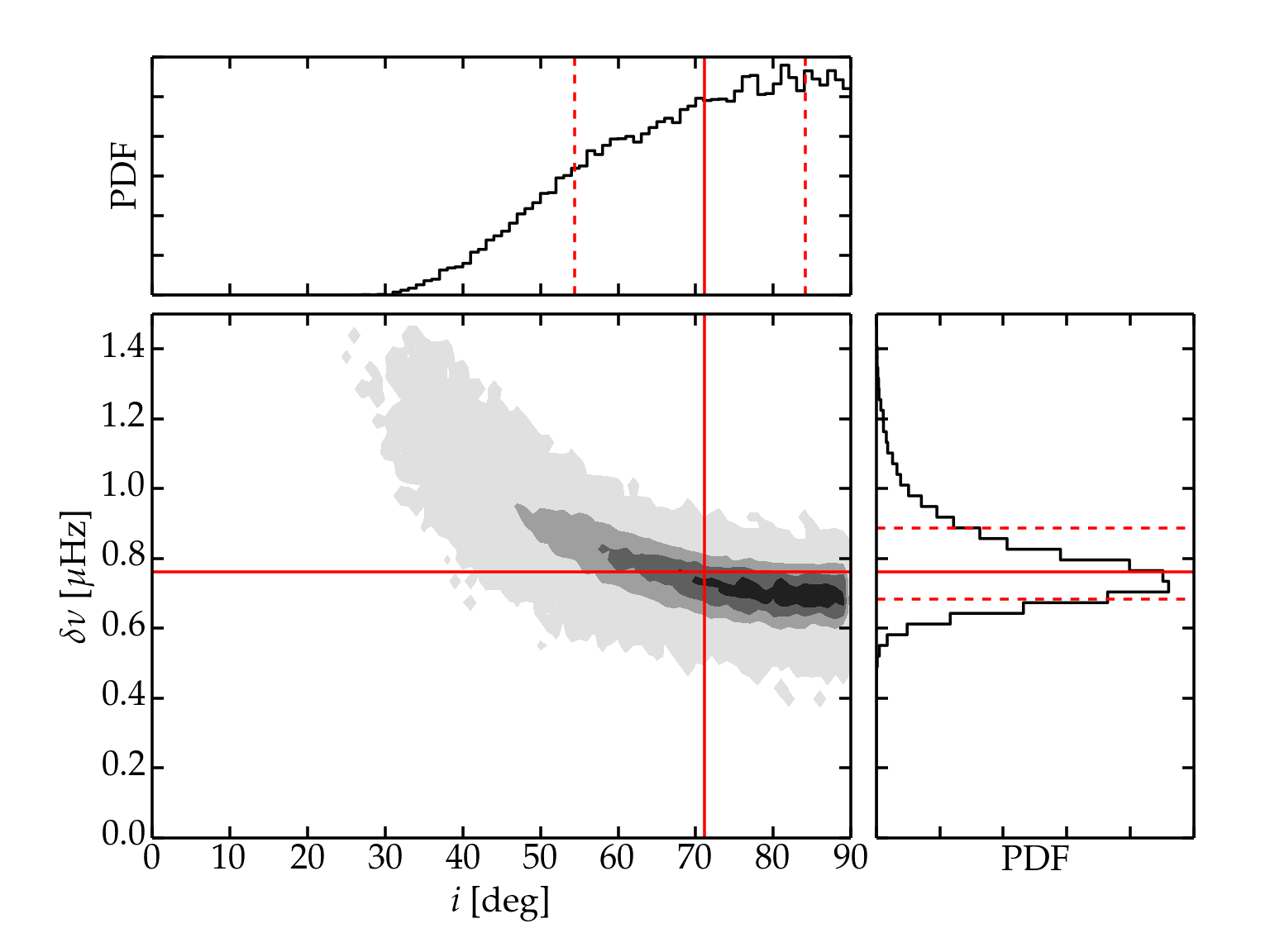}
		\caption{Bottom left: A 2D representation of the marginalized posterior distributions for the rotational splitting $\delta \nu$ and the inclination of the rotation axis $i$. Top and right frames show the projection onto each axis in solid black. The solid red lines indicate the median along each axis, and dashed red lines are $16th$, $84th$ percentile values. These distributions are obtained from the local fit to the modes shown in Fig. \ref{fig:perky_modelfig}.}
	\label{fig:perky_posteriors}
\end{figure}

The complete list of fit values and associated errors (Table 2) is available as online material via the CDS. The fit values for the rotational splitting $\delta \nu$ and the inclination angle $i$ are presented in Fig.~\ref{fig:split_and_incl}. We compute a variance weighted mean of the splittings measured for each star, and list these in Table~\ref{tab:results}. The posterior distributions of the rotational splittings are approximately Gaussian around the mean (see Fig. \ref{fig:perky_posteriors}), so the variance is representative of the errors associated with each splitting. This is not true for the posterior distributions of the inclinations and so we cannot apply this to obtain a weighted mean value representative of the inclination of each star. We therefore only list an unweighted mean of the inclination measurements with typical errors of $\sim 20^{\rm{o}}$.

A few stars (e.g., \object{KIC006106415}) appear to show a marginal trend in the splittings with increasing frequency. To test this further we computed a $\chi^2$ and the associated p-values based on the variance weighted mean splitting, i.e., a constant rotational splitting with frequency. We found that the $\chi^2$ values ranged between $0.6-3.9$ and the p-values between $0.69-0.998$, indicating that the measurements are consistent with a constant splitting over frequency. We noted that the errors on the rotational splittings are likely to be anti-correlated with the errors on the inclinations (see Fig. \ref{fig:perky_posteriors}). We therefore also computed posterior distributions of $\delta \nu \sin{i}$ (middle row in Fig. \ref{fig:split_and_incl}), and performed the same test for constant rotation. The computed $\chi^2$ and p-values were between $3.3-7.7$ and $0.36-0.77$, respectively, i.e., the variations seen in $\delta \nu \sin{i}$ are still consistent with uniform rotation in these stars. We therefore find no evidence of differential rotation in these stars. In Sun-like stars the mode linewidths increase strongly with frequency \citep{Chaplin1998}. This means that using a common linewidth likely ceases to be a good approximation for the last few mode sets at higher frequencies, thus introducing a bias in the splitting parameter. 
\begin{table*}
\centering
\caption{Variance weighted mean rotational splittings $\langle \delta \nu \rangle$, inclination $\langle i \rangle$,  and rotation period $\Omega/\Omega_\odot$ for the six Sun-like stars. The effective temperature $T_{\rm{eff}}$, surface gravity $\log{g}$, and frequency intervals considered for each star are also listed, where each interval is divided into segments of length approximately equal to the large frequency separation. The variance weighted mean splittings $\langle \delta \nu \rangle$ are shown as dashed lines in Fig. \ref{fig:split_and_incl}, where the listed errors are the standard deviations of the weighted mean values. We note that the posterior distributions for the $\delta \nu$ are only approximately Gaussian. The posterior distributions of inclination measurements cannot be approximated as a Gaussian and so we only show the unweighted mean of the inclinations where typical errors are $\sim 20^{\rm{o}}$. The reader should not use the mean values and associated errors reported here, but should refer to the online material for more accurate values for each mode set. For comparison, the final column shows the stellar rotation rate relative to the solar value (we used $\Omega_\odot = 0.424\mu$Hz).}  

\begin{tabular}{llcccccc}
\toprule
Star         & $T_{\rm{eff}}$ [K]         & $\log{g}$ [$\rm{cm/s}^2$]         & Fit interval [$\mu$Hz] & $\langle \delta \nu \rangle$ [$\mu$Hz]              & $\langle \delta \nu \sin{i} \rangle$ [$\mu$Hz]& $\langle i \rangle$ [deg]         & $\Omega/\Omega_{\odot}$  \\
\midrule
\object{KIC004914923} & $5808\pm92$  & $4.28\pm0.21$ & 1429 - 2135 & $0.522\pm 0.074$&$0.371\pm 0.029$ & $54$ & $1.23 \pm 0.29$\\
\object{KIC005184732} & $5669\pm97$  & $4.07\pm0.21$ & 1632 - 2400 & $0.643\pm 0.063$&$0.517\pm 0.027$ & $62$ & $1.52 \pm 0.12$\\
\object{KIC006106415} & $6050\pm70$  & $4.40\pm0.08$ & 1677 - 2609 & $0.708\pm 0.038$&$0.647\pm 0.022$ & $64$ & $1.67 \pm 0.27$\\
\object{KIC006116048} & $5991\pm124$ & $4.09\pm0.21$ & 1620 - 2425 & $0.703\pm 0.053$&$0.603\pm 0.024$ & $69$ & $1.66 \pm 0.36$\\
\object{KIC006933899} & $5837\pm97$  & $4.21\pm0.22$ & 1157 - 1662 & $0.404\pm 0.078$&$0.296\pm 0.034$ & $57$ & $0.95 \pm 0.27$\\
\object{KIC010963065} & $6097\pm130$ & $4.00\pm0.21$ & 1760 - 2475 & $0.801\pm 0.079$&$0.656\pm 0.032$ & $56$ & $1.89 \pm 0.20$
\end{tabular}
\label{tab:results}
\end{table*}

The inclination of the rotation axis is an important parameter for characterizing exoplanetary systems and constraining models of planet formation and evolution (e.g., Nagasawa et al. 2008). However, we found that the inclination angles are very poorly constrained when using a single mode set, even with these prime examples from the \textit{Kepler} database. In Fig.~\ref{fig:perky_posteriors} we show the marginalized posterior distributions for the fit shown in Fig.~\ref{fig:perky_modelfig}. The posterior distribution reveals that the inclination angle is dominated by the $\sin i$ prior, i.e., an individual mode set yields very little information about the stellar inclination axis. In this case, based on the posterior distribution we could only conclude that $i\lesssim45^{\rm{o}}$ is unlikely. This is a common trait of the posterior distributions for the other stars in our sample, and some are even less constrained so that we can only rule out $i\lesssim20^{\rm{o}}$. The relatively high inclination angles that we measure are expected when considering these stars were chosen by eye to have a visible splitting, or at least a broadening of the $l=2$ and $l=1$ modes. This selection naturally biases the sample of stars toward highly inclined configurations \citep[see Fig. 2. in][]{Gizon2003}.

These stars were specifically selected for this study since they have visible rotational splittings. When using high signal-to-noise observations such as these, it is a simple matter of fitting just the central mode sets of the $p$-mode envelope in order to obtain a reliable measure of the rotational splitting. Furthermore, these high-quality data offer the tantalizing possibility of measuring radial differential rotation.  From our measurements we have determined that these Sun-like stars are unlikely to have variations in rotational splittings larger than $\sim 40\%$. Improvements to the fitting method, e.g., linewidth parametrization or a global fit to the power spectrum, could reduce the uncertainties on the splitting measurements and potentially reveal the signatures of differential rotation.

\begin{acknowledgements}
M.N., L.G. and H.S. acknowledge research funding by Deutsche Forschungsgemeinschaft (DFG) under grant SFB 963/1 ``Astrophysical flow instabilities and turbulence'' (Project A18, WP ``Seismology of magnetic activity''). This paper includes data collected by the \textit{Kepler} mission. Funding for the \textit{Kepler} mission is provided by the NASA Science Mission directorate. Data presented in this paper were obtained from the Mikulski Archive for Space Telescopes (MAST). STScI is operated by the Association of Universities for Research in Astronomy, Inc., under NASA contract NAS5-26555. Support for MAST for non-HST data is provided by the NASA Office of Space Science via grant NNX09AF08G and by other grants and contracts. 
\end{acknowledgements}

\bibliographystyle{aa}
\bibliography{peakbagging_rotation_paper}

\end{document}